\documentclass{article}
\usepackage{spconf,amsmath,graphicx,amssymb}
\usepackage{multirow}
\usepackage{subfigure}
\usepackage{stfloats}

\title{Video-based Person Re-identification with Two-stream Convolutional Network and Co-attentive Snippet Embedding}
\name{Peixian Chen, Pingyang Dai, Qiong Wu, Yuyu Huang}
\address{School of Information Science and Engineering, Xiamen University \\ 
\texttt{\{pxchen, qiong, huangyuyu\}@stu.xmu.edu.cn, pydai@xmu.edu.cn}}

\begin{document}
%
\maketitle
\begin{abstract}
Recently, the applications of person re-identification in visual surveillance and human-computer interaction are sharply increasing, which signifies the critical role of such a problem. In this paper, we propose a two-stream convolutional network (ConvNet) based on the \emph{competitive similarity aggregation scheme} and \emph{co-attentive embedding strategy} for video-based person re-identification. By dividing the long video sequence into multiple short video snippets, we manage to utilize every snippet's RGB frames, optical flow maps and pose maps to facilitate residual networks, e.g., ResNet, for feature extraction in the two-stream ConvNet. The extracted features are embedded by the \emph{co-attentive embedding} method, which allows for the reduction of the effects of noisy frames. Finally, we fuse the outputs of both streams as the embedding of a snippet, and apply \emph{competitive snippet-similarity aggregation} to measure the similarity between two sequences. Our experiments show that the proposed method significantly outperforms current state-of-the-art approaches on multiple datasets.
\end{abstract}
\section{Introduction}
\label{sec:intro}
Person re-identification is a problem of re-matching pedestrians on non-overlapping cameras \cite{Cheng2016Person}. It can be applied to video surveillance and has experienced full attention in recent years. It can be regarded as a generic object recognition task. Previous methods in the literature mainly focus on the image-based settings, which compare images of one person with others captured by different cameras \cite{Cheng2016Person, Fei2014Person, Zhao2017Person, Farenzena2010Person}. 

Video-based person re-identification applications, however, are believed to be more valuable and robust, as video data contains richer information concerning the appearance of pedestrians and conveys action clues that implicitly reflect the layout of the human body as well. Subsequently, some works have been explored to investigate video-to-video matching problems in person re-identification \cite{Xu2017Jointly, wang2014person, Zhao2017Person,chen2018video}. 

Existing video-based person re-identification methods can be classified into two categories: metric learning and video data representation. For methods based on metric learning, the primary goal is to reduce the intra-class variance, such as learning a dictionary to sparsely encode features \cite{karanam2015person}, learning the top-push distance \cite{you2016top}, and ranking and selecting video segments \cite{wang2014person}. As for methods based on video data representation, traditional ones typically incorporate temporal information \cite{karanam2015sparse, liu2015spatio, chen2016person}. 

Among those methods, an intuitive way to represent video data is to feed the whole video sequence into the network that outputs an embedding vector.  Such kind of representations, however, can preserve neither critical details that facilitate recognition nor temporal information that helps to distinguish similar walking postures, as video sequences can bear considerable visual variations. Though this problem has been extensively explored in the literature, there remain to be several challenges due to the low quality of videos and the similarity of the pedestrian's actions.

\begin{figure*}[hb]
    \centering
    \includegraphics[width=0.7\linewidth]{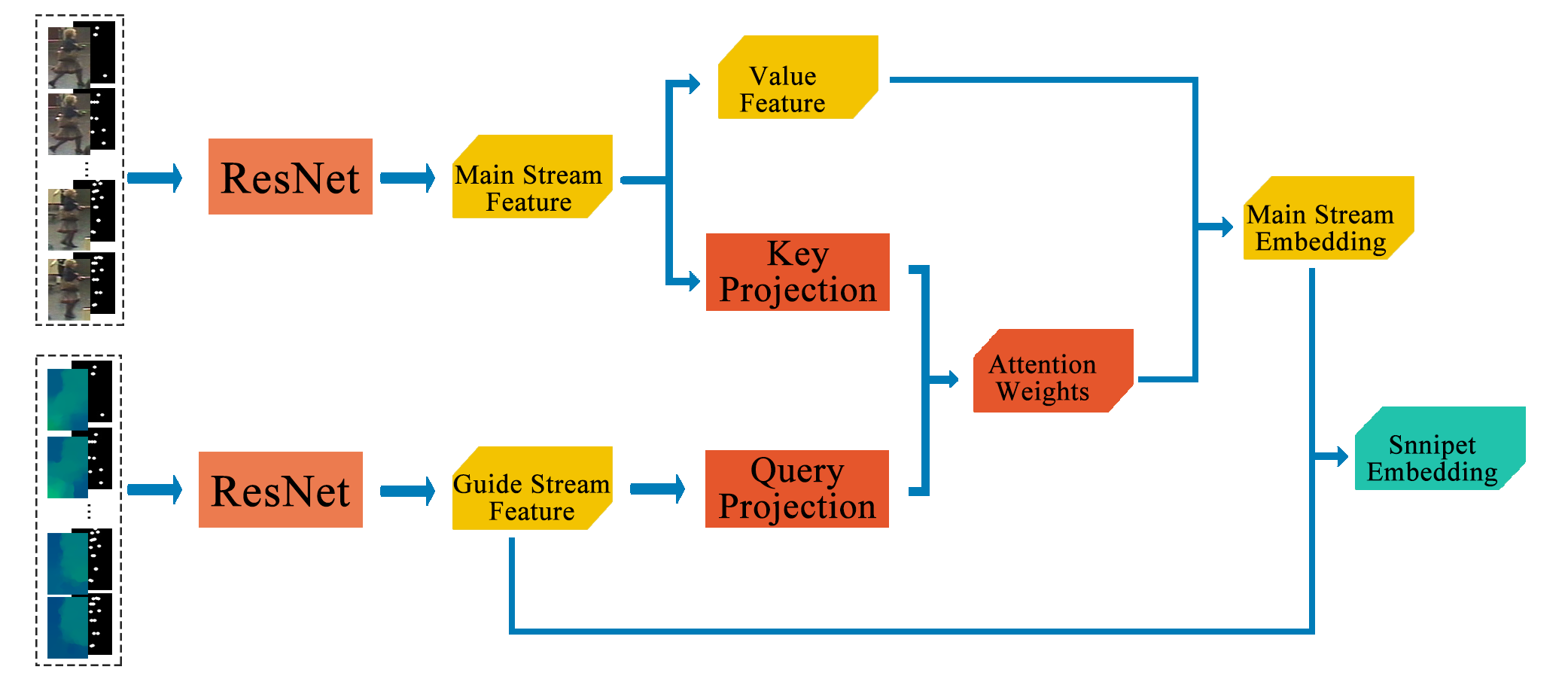}
    \caption{Our video-based person re-identification system is constructed based on a two-stream ConvNet \cite{Simonyan2014Two}. The inputs of the first stream, i.e., \emph{main} stream, are RGB frames and pose maps; as for the second stream, i.e., \emph{guide} stream, its inputs are pose and optical maps. They are directed into a CNN network, and the subsequent features are processed through the co-attentive embedding. Finally, the \emph{main} stream's embedding vector and the \emph{guide} stream's extracted feature are combined and subsequently generate the snippet embedding.}
    \label{fig:network}
\end{figure*}

In this paper, we propose our method based on a two-stream ConvNet (Section \ref{sec:two-stream}), which divides a video sequence into multiple snippets that are later embedded with \emph{co-attentive embedding mechanism} (Section \ref{sec:coattention}), and measure the similarity between two long sequences with \emph{competitive snippet-similarity aggregation scheme} (Section \ref{sec:snippet-similarity}). Figure \ref{fig:network} illustrates the basic idea of out approach.

Our main contributions are summarized as follows. (1) We apply the two-stream ConvNet to person re-identification and incorporate the \emph{competitive snippet-similarity aggregation scheme} and \emph{co-attentive embbedding mechanism}. This network architecture reduces the variations in the appearance of the segment similarity measurement and the influence of noise frames, which result in better embedding features for similarity measurement. (2) Apart from the RGB frame and optical flow maps being used in the above methods, we introduce pose maps to this problem as well. This modification can provide additional pedestrian's gait information and a more accurate position of the person, which is more effective in obtaining important features for person recognition. (3) We evaluate our approach on two person re-identification datasets (iLIDS-VID and PRID2011), the experimental results demonstrate that our approach outperforms state-of-the-art methods with top-1 accuracy of 88.7 \% on the iLIDS-VID dataset and 94.4\% on the PRID2011 dataset.

\section{APPROACH}
\label{sec:approach}

We consider video-based person re-ID as a ranking problem, as its objective is retrieving sequences of the same person from a gallery set according to a probe person sequence, which is typically solved by sorting the gallery set based on the similarity to the probe sequence and output the most similar one. In this paper, we propose the scaled dot-product attention \cite{attention} for video-based person re-ID based on a two-stream network architecture \cite{Simonyan2014Two}. Moreover, according to the method proposed by \cite{chen2018video}, we break long sequences into short snippets as to measure the similarity between snippets and later aggregate the top-ranked snippet competitively.

\subsection{Snippet-similarity Aggregation}
\label{sec:snippet-similarity}
Let $p$ and $g$ denote a probe sequence and a gallery sequence, respectively. We consider our goal as to measure the similarity between $p$ and each $g$ in the gallery set and choose the most similar one. Instead of comparing two sequences directly, we measure the similarity between two sequences with a snippet-wise method. 

Specifically, we extract $N_s$ snippets from each sequence by sampling $L$ from every $T$ frames in the sequence of each snippet. Let $S_s$ denote the set of snippets sampled from sequence $s$ with $F_s$ frames; we determine the value of $N_s$ as follows. If there are sufficient frames, i.e., $F_s$ is no less than $L$, $N_s$ is given by Equation \ref{equ:numofsnippet} with the last several frames discarded. Otherwise, when $F_s$ is smaller, there will be only one snippet containing all frames with the last one replicated until the designated snippet length is satisfied.

\begin{equation}
N_s = \left\lfloor \frac{F_s-1-L}{T}+1 \right\rfloor
\label{equ:numofsnippet}
\end{equation}

All snippets sampled from a particular sequence $s$ form a snippet set $\mathcal{S}^s$ of size $N_s$. We therefore measure the similarity between each pair of snippets $(p_i, g_j)$ from two snippet sets, namely $\mathcal{S}^p$ for the probe sequence and $\mathcal{S}^g$ for the gallery sequence, which results in a $N_p\times N_g$ score matrix $D$ as described in Equation \ref{eq:similarity}, where $p_i$ and $g_j$ are two arbitrary snippets in $\mathcal{S}^p$ and $\mathcal{S}^g$, respectively.

Particularly, we define the similarity between two snippets with a CNN-based distance metric $d(\cdot)$. Note that in a distance metric, a lower score indicates a higher similarity.

\begin{equation}
D(p, g) = \big\{d(p_i, g_j) \mathop{\big|} p_i\in \mathcal{S}^p, g_j \in \mathcal{S}^g\big\}
\label{eq:similarity}
\end{equation}

To resolve the error caused by visual variations, we employ a competitive strategy \cite{chen2018video}, which computes an aggregated similarity metric $\hat{D}$ based on top-\emph{k} ranked snippet-similarity scores, and thus defines the ultimate similarity $s(p, g)$ between two sequences $p$ and $g$ as the average of all scores indicated by Equation \ref{eq:aggregation}.

\begin{equation}
s(p,g) = \sum_{\hat{d}\in\hat{D}(p,g)}\hat{d}\bigg/\left|\hat{D}(p, g)\right|
\label{eq:aggregation}
\end{equation}

\subsection{Two-stream Architecture}
\label{sec:two-stream}

We assume that a video contains both spatial and temporal information, among which the spatial one indicates visual appearance feature of intra-frames, such as the scene and object depicted in the video; whereas the temporal one conveys the inter-frame feature, such as the motion of objects. Notably, we also assume that pedestrian's gait information could provide rich information that contributes to the problem of video-based person re-identification.

As a result, we devise our network architecture accordingly by dividing it into two streams. Both streams are implemented based on the ImageNet pre-trained \emph{ResNet-50} model, and the two subsequent features are processed through a co-attentive embedding mechanism and combined by late fusion.

We name the two streams as the \emph{main} stream and \emph{guide} stream based on their desired responsibilities. The input of the \emph{main} stream is a channel-wise concatenation of the RGB frame and pose maps, where the former indicates visual information while the latter provides both pedestrian's position and gait information, thus facilitate recognition of the person's area. Whereas in the \emph{guide} stream, the inputs are the pose and optical flow maps. We then use an attention network, which will be discussed in more details in Section \ref{sec:coattention}, to process the outputs of both streams, which lead to a query and a set of key-value pairs, respectively.

The network structure described above is constructed with a co-attentive mechanism, such that both the pose and optical flow maps can perform as weights of the features extracted from the RGB frame, and thus embed a video snippet into a single vector. Furthermore, we combine the outputs of both \emph{guide} stream and attention network as the embedding vector of a snippet and measure the similarity between two snippets based on the distance of their embedding vectors.

\subsection{Co-attention Snippet Embedding}
\label{sec:coattention}
To measure the similarity between two snippets, we propose a co-attentive mechanism based on the scaled dot-product attention \cite{attention}. In both streams, we employ the visual CNN to extract per-frame features from each of the $L$ frames of snippet $s$. Denote the feature of all frames as ${\boldsymbol\chi_L}$ and $l$-th frame’s feature vector as $\chi_l$.

Since visual features of all frames are contained in the resulting feature $\boldsymbol{\chi_L}$, we assume the existance of redundant information in such features. We then propose an attention mechanism to distill the information of the sequential features with the guidance of both pose and optical flow maps. Specifically, we generate a query feature for each snippet and a key-value feature for each frame based on the features extracted from both streams, respectively. The ultimate snippet's embedding vector is a weighted summation of the feature of all frames, where the weight is determined by the compatibility of the query features $\mathfrak{q}$ and key feature $\mathfrak{K}$. 

We then use a fully connected layer and a BN layer to construct a query and a key projections based on the extracted features, which are also used as the value feature $\mathfrak{V}$ for each frame. We average the output of query fully connected layer of each frames for each snippet as the query feature $\mathfrak{q}$. After that, the dot-product of $\mathfrak{q}$ and $\mathfrak{K}$ are computed with softmax non-linearization to generate the attention weights $\mathfrak{w}$ for each snippet. Finally, we compute the sum of the per-frame value feature weighted by the attention, which results in the embedding feature of snippet $s$ given by $\mathcal{F} = \mathfrak{w}\cdot\mathfrak{V}$.



\section{EXPERIMENTS}
\label{sec:experiments}

\begin{table*}[h!]
\centering
\label{tab:result_ilids_prid}

\begin{tabular}{c|cccc|cccc}
\hline
\multirow{2}{*}{Methods} & \multicolumn{4}{c|}{\textbf{iLIDS-VID}} & \multicolumn{4}{c}{\textbf{PRID2011}} \\ \cline{2-9} 
 & Top-1 & Top-5 & Top-10 & Top-20 & Top-1 & Top-5 & Top-10 & Top-20 \\ \hline\hline
Karanam et al. \cite{karanam2015person} & 25.9 & 48.2 & 57.3 & 68.9 & 40.6 & 69.7 & 77.8 & 85.6 \\
Wang et al.\cite{wang2014person} & 41.3 & 63.5 & 72.7 & 83.1 & 48.3 & 74.9 & 87.3 & 94.4 \\
Cho et al. \cite{cho2016improving}& 30.3 & 56.3 & 70.3 & 82.7 & 45.0 & 72.0 & 85.0 & 92.5 \\
You et al.\cite{you2016top} & 56.3 & 87.6 & 95.6 & 98.3 & 56.7 & 80.0 & 87.6 & 93.6 \\
McLaughlin et al. \cite{mclaughlin2016recurrent} & 58.0 & 84.0 & 91.0 & 96.0 & 70.0 & 90.0 & 95.0 & 97.0 \\
Zhang et al. \cite{zhang2017attributes} & 60.3 & 85.3 & 93.6 & 98 & 73.2 & 93.0 & 96.3 & 98.3 \\
Zheng et al. \cite{zheng2016mars} & 53.0 & 81.4 & -- & 95.1 & 77.3 & 93.5 & -- & 99.3 \\
Zhou et al. \cite{zhou2017see} & 55.2 & 86.5 & -- & 97.0 & 79.4 & 94.4 & -- & 99.3 \\
Chen et al. \cite{chen2018video} & 85.4 & 96.7 & 98.8 & 99.5 & 93.0 & 99.3 & 100.0 & 100.0 \\ \hline
Ours & \textbf{88.7} & \textbf{98} & \textbf{99.3} & \textbf{100.0} & \textbf{94.4} & \textbf{99.3} & \textbf{100.0} & \textbf{100.0} \\ \hline
\end{tabular}
\caption{Top-\emph{k} accuracy (\%) of existing methods on iLIDS-VID and PRID2011 datasets.}
\end{table*}

\subsection{Datasets and setting}
We evaluate our proposed method on two datasets, iLIDS-VID \cite{wang2014person} and PRID-2011 \cite{Hirzer2011Person}. The former contains 600 videos of 300 persons, which have variable lengths from 23 to 193 frames. We randomly split the dataset by half based on the person's identification for the training and testing sets. As for the latter, it includes 749 persons and 400 videos with 5 to 675 frames per video. We then choose 178 persons and divide them into the training and testing set by half like  \cite{wang2014person}.

We then use the average Cumulative Matching Characteristic (CMC) curve and Mean Average Precision (\emph{m}AP) as the measurement indicator. The length $L$ of sampled snippets and step size $T$ are designated as 8 and 3, respectively.

\subsection{Result and anlaysis}
\label{sec:result}
We compare our method with most of the existing approaches and present our results in Table \ref{tab:result_ilids_prid}. Note that these results are achieved without using any post-processing techniques, such as re-ranking \cite{ZhongZCL17} and multi-query \cite{zheng2016mars}. 

\textbf{Compared with state-of-the-art methods} As shown in Table \ref{tab:result_ilids_prid}, our method is 30\%-60\% and 38\%-54\% higher on top-1 accuracy than those based on handcrafted features \cite{karanam2015person, wang2014person, cho2016improving, you2016top} on the two datasets respectively. 

Recently, some methods have applied deep learning frameworks to solve the problem of person re-identification, such as LSTM \cite{mclaughlin2016recurrent}, pooling \cite{zheng2016mars}, and temporal attention \cite{zhou2017see} for feature summarization \cite{chen2018video}. Some of these method neglect temporal information \cite{zhang2017attributes}, while some others embed individual images or the whole sequence into a vector \cite{mclaughlin2016recurrent,you2016top, zheng2016mars, zhou2017see}. 

Our approach mainly improves the method proposed by \cite{chen2018video}, which splits the sequence into video snippets for both training and testing set. In addition, we also employ a two-stream ConvNet with co-attention and additional pose information, which makes the embedded snippets more relevant to the probe snippet for similarity measurement. Finally, our experiments demonstrate that we have further improved the performance achieved by \cite{chen2018video}. 


\begin{figure*}
\centering
\subfigure[Top-1 accuracy and \emph{m}AP score on the iLIDS-VID dataset]{
    \label{fig:ilidresult}
    \includegraphics[width=0.48\linewidth]{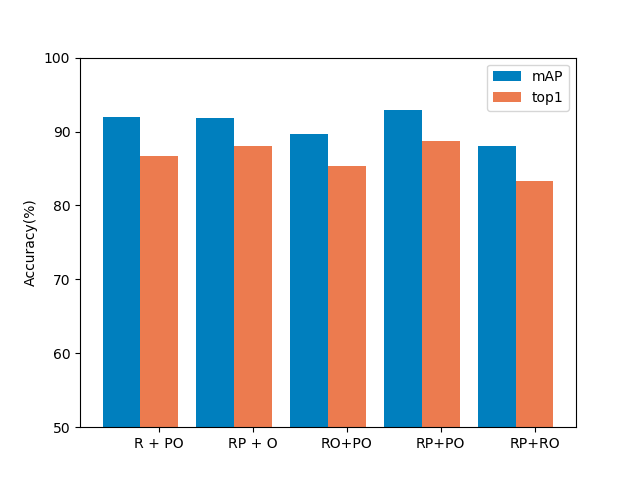}
}
\subfigure[Top-1 accuracy and \emph{m}AP score on the PRID2011 dataset]{
    \label{fig:pridresult}
    \includegraphics[width=0.48\linewidth]{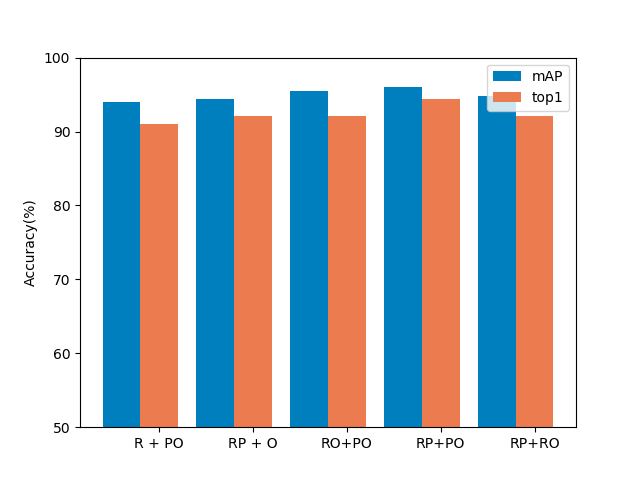}
}
\caption{Parameter analysis for inputs of the \emph{main} stream and \emph{guide} stream. Figure \ref{fig:ilidresult} and \ref{fig:pridresult} explain the top-1 accuracy and \emph{m}AP score on iLIDS-VID and PRID2011 datasets with 5 different inputs. Note that \textbf{R}, \textbf{O}, and \textbf{P} denote the RGB frame, optical flow and pose maps, respectively. Inputs of the \emph{main} and \emph{guide} streams are designated before and after the $+$ sign.}
\label{fig:ilidsvidresult}
\end{figure*}


\textbf{The influence of the inputs of two-stream ConvNet} To demonstrate the impact of different inputs on the accuracy of the two-stream ConvNet, we conduct a series of experiments using different combinations of inputs of the two-steam network in the iLIDS-VID and PRID2011 datasets. Figure \ref{fig:ilidresult} and \ref{fig:pridresult} illustrate the top-1 accuracy and mAP on both datasets with different combinations of inputs. The result demonstrates that our proposed network is more effective under various inputs. All the top-1 accuracies are above 83\% and \emph{m}APs are above 88\%. Taking RGB frames and pose maps as inputs of the \emph{main} stream and pose maps and optical flow as inputs of the \emph{guide} stream can be more effective than others on both iLIDS-VID and PRID2011 datasets. 

We recognize the explanation as that the RGB frames contain spatial information, and pose maps benefit capturing the person's positions and their following combinations, which have the capability of describing more extensive frame information. The human gait information within pose maps also plays an essential role, which means its combination with optical flow maps would contribute to better attention weights.

Although the above settings, in which the \emph{main} stream contains RGB frames and pose maps and the \emph{guide} stream contains both pose and optical flow maps, could obtain satisfied accuracy, we find that removing the pose maps from the \emph{guide} stream can considerably reduce the cost of time and computation resources while sacrificing acceptable accuracy.

\section{CONCLUTION}
\label{sec:conclution}
In this paper, we propose a two-stream ConvNet with the competitive similarity aggregation scheme and co-attentive embedding mechanism for video-based person re-identification, and take RGB frames, optical flow maps and pose maps as inputs to utilize the rich information that a video contains.

Such architecture not only observes the person's appearance information, spatial and temporal information; but also incorporates pedestrian's gait information. This architecture can reduce intra-frame variations in each sampled snippet, alleviate the impacts of noisy frames, and enforce the compared snippet pairs weighting more on related information for snippet-similarity estimation. We evaluate our approach on two datasets and the experiment results demonstrate that our method outperforms other state-of-the-art techniques.

\bibliographystyle{IEEEbib}
\bibliography{refs}

\begin{thebibliography}{10}

\bibitem{Cheng2016Person}
De~Cheng, Yihong Gong, Sanping Zhou, Jinjun Wang, and Nanning Zheng,
\newblock ``Person re-identification by multi-channel parts-based cnn with
  improved triplet loss function,''
\newblock in {\em Computer Vision \& Pattern Recognition}, 2016.

\bibitem{Fei2014Person}
Xiong Fei, Mengran Gou, Octavia Camps, and Mario Sznaier,
\newblock ``Person re-identification using kernel-based metric learning
  methods,''
\newblock in {\em European Conference on Computer Vision}, 2014.

\bibitem{Zhao2017Person}
R.~Zhao, W.~Ouyang, and X.~Wang,
\newblock ``Person re-identification by saliency learning,''
\newblock {\em IEEE Transactions on Pattern Analysis \& Machine Intelligence},
  vol. 39, no. 2, pp. 356--370, 2017.

\bibitem{Farenzena2010Person}
M.~Farenzena, L.~Bazzani, A.~Perina, V.~Murino, and M.~Cristani,
\newblock ``Person re-identification by symmetry-driven accumulation of local
  features,''
\newblock in {\em Computer Vision \& Pattern Recognition}, 2010.

\bibitem{Xu2017Jointly}
Shuangjie Xu, Yu~Cheng, Kang Gu, Yang Yang, Shiyu Chang, and Pan Zhou,
\newblock ``Jointly attentive spatial-temporal pooling networks for video-based
  person re-identification,''
\newblock {\em arXiv preprint arXiv:1708.02286}, 2017.

\bibitem{wang2014person}
Taiqing Wang, Shaogang Gong, Xiatian Zhu, and Shengjin Wang,
\newblock ``Person re-identification by video ranking,''
\newblock in {\em European Conference on Computer Vision}. Springer, 2014, pp.
  688--703.

\bibitem{chen2018video}
Dapeng Chen, Hongsheng Li, Tong Xiao, Shuai Yi, and Xiaogang Wang,
\newblock ``Video person re-identification with competitive snippet-similarity
  aggregation and co-attentive snippet embedding,''
\newblock in {\em Proceedings of the IEEE Conference on Computer Vision and
  Pattern Recognition}, 2018, pp. 1169--1178.

\bibitem{karanam2015person}
Srikrishna Karanam, Yang Li, and Richard~J Radke,
\newblock ``Person re-identification with discriminatively trained viewpoint
  invariant dictionaries,''
\newblock in {\em Proceedings of the IEEE International Conference on Computer
  Vision}, 2015, pp. 4516--4524.

\bibitem{you2016top}
Jinjie You, Ancong Wu, Xiang Li, and Wei-Shi Zheng,
\newblock ``Top-push video-based person re-identification,''
\newblock in {\em Proceedings of the IEEE Conference on Computer Vision and
  Pattern Recognition}, 2016, pp. 1345--1353.

\bibitem{karanam2015sparse}
Srikrishna Karanam, Yang Li, and Richard~J Radke,
\newblock ``Sparse re-id: Block sparsity for person re-identification,''
\newblock in {\em Proceedings of the IEEE conference on computer vision and
  pattern recognition workshops}, 2015, pp. 33--40.

\bibitem{liu2015spatio}
Kan Liu, Bingpeng Ma, Wei Zhang, and Rui Huang,
\newblock ``A spatio-temporal appearance representation for viceo-based
  pedestrian re-identification,''
\newblock in {\em Proceedings of the IEEE International Conference on Computer
  Vision}, 2015, pp. 3810--3818.

\bibitem{chen2016person}
Jiaxin Chen, Yunhong Wang, and Yuan~Yan Tang,
\newblock ``Person re-identification by exploiting spatio-temporal cues and
  multi-view metric learning,''
\newblock {\em IEEE Signal Processing Letters}, vol. 23, no. 7, pp. 998--1002,
  2016.

\bibitem{Simonyan2014Two}
Karen Simonyan and Andrew Zisserman,
\newblock ``Two-stream convolutional networks for action recognition in
  videos,''
\newblock 2014.

\bibitem{attention}
Ashish Vaswani, Noam Shazeer, Niki Parmar, Jakob Uszkoreit, Llion Jones,
  Aidan~N Gomez, {\L}ukasz Kaiser, and Illia Polosukhin,
\newblock ``Attention is all you need,''
\newblock in {\em Advances in Neural Information Processing Systems}, 2017, pp.
  5998--6008.

\bibitem{cho2016improving}
Yeong-Jun Cho and Kuk-Jin Yoon,
\newblock ``Improving person re-identification via pose-aware multi-shot
  matching,''
\newblock in {\em Proceedings of the IEEE conference on computer vision and
  pattern recognition}, 2016, pp. 1354--1362.

\bibitem{mclaughlin2016recurrent}
Niall McLaughlin, Jesus Martinez~del Rincon, and Paul Miller,
\newblock ``Recurrent convolutional network for video-based person
  re-identification,''
\newblock in {\em Proceedings of the IEEE conference on computer vision and
  pattern recognition}, 2016, pp. 1325--1334.

\bibitem{zhang2017attributes}
Xiu Zhang, Federico Pala, and Bir Bhanu,
\newblock ``Attributes co-occurrence pattern mining for video-based person
  re-identification,''
\newblock in {\em Advanced Video and Signal Based Surveillance (AVSS), 2017
  14th IEEE International Conference on}. IEEE, 2017, pp. 1--6.

\bibitem{zheng2016mars}
Liang Zheng, Zhi Bie, Yifan Sun, Jingdong Wang, Chi Su, Shengjin Wang, and
  Qi~Tian,
\newblock ``Mars: A video benchmark for large-scale person re-identification,''
\newblock in {\em European Conference on Computer Vision}. Springer, 2016, pp.
  868--884.

\bibitem{zhou2017see}
Zhen Zhou, Yan Huang, Wei Wang, Liang Wang, and Tieniu Tan,
\newblock ``See the forest for the trees: Joint spatial and temporal recurrent
  neural networks for video-based person re-identification,''
\newblock in {\em Computer Vision and Pattern Recognition (CVPR), 2017 IEEE
  Conference on}. IEEE, 2017, pp. 6776--6785.

\bibitem{Hirzer2011Person}
Martin Hirzer, Csaba Beleznai, Peter~M. Roth, and Horst Bischof,
\newblock ``Person re-identification by descriptive and discriminative
  classification,''
\newblock in {\em Scandinavian Conference on Image Analysis}, 2011.

\bibitem{ZhongZCL17}
Zhun Zhong, Liang Zheng, Donglin Cao, and Shaozi Li,
\newblock ``Re-ranking person re-identification with k-reciprocal encoding,''
\newblock {\em CoRR}, vol. abs/1701.08398, 2017.

\end{thebibliography}

\end{document}